\documentstyle[11pt,epsf]{article}
\newcommand{\bce}{\begin{center}}
\newcommand{\ece}{\end{center}}
\newcommand{\be}{\begin{equation}}
\newcommand{\ee}{\end{equation}}
\newcommand{\bea}{\begin{eqnarray}}
\newcommand{\eea}{\end{eqnarray}}
\newcommand{\E}{\>=\>}
\newcommand{\EA}{&=&}

\begin{document}

\setcounter{page}{0}
\thispagestyle{empty}

\vspace{2.5cm}

\bce
{\large\bf On the Best Quadratic Approximation in Feynman's}

\vspace{0.3cm}
{\large\bf Path Integral Treatment of the Polaron}

\vspace{2cm}
R. Rosenfelder $^1$ and A. W. Schreiber $^2$

\vspace{0.8cm}
$^1$ Paul Scherrer Institute, CH-5232 Villigen PSI, Switzerland

\vspace{0.1cm}
$^2$ Department of Physics and Mathematical Physics, and
           Research Centre for the Subatomic Structure of Matter,
           University of Adelaide, Adelaide, S. A. 5005, Australia

\ece

\vspace{4cm}

\begin{abstract}
\noindent
The best quadratic approximation to the retarded polaron action due
to Adamowski {\it et al.} and Saitoh is investigated numerically for a wide
range of coupling constants. The non-linear variational equations are 
solved iteratively with an efficient method in order to obtain 
the ground-state energy and the effective mass of the polaron. The virial 
theorem and expansions for small and large couplings are used to check the 
high accuracy of the numerical results.
Only small improvements over Feynman's (non-optimal) results are observed.
For a moving polaron it is shown that the most general quadratic 
trial action may contain anisotropic terms which, however, do not lead to 
improvements for the ground-state energy and effective mass.
\end{abstract}

\vspace{2cm}
PACS number(s): 71.38.Fp, 03.65.Db \\


\newpage
{\bf 1.} The polaron problem is a non-relativistic field theory for an 
electron moving in a crystal and has received a lot of attention in the 
past decades (for reviews see \cite{GeLo}, \cite{MCM}). Among the many 
theoretical treatments Feynman's approach \cite{Fey} is still outstanding: 
he first integrated out the phonons to obtain an effective action for the 
electron which for large Euclidean times reads
\be
S_{\rm \> eff} \E \int_0^{\beta} dt \> \frac{1}{2}{\bf \dot x}^2 -
\frac{\alpha}{2 {\sqrt 2}} \int_0^{\beta} dt \, dt' \> \frac{e^{-(t-t')}}
{|\> {\bf x}(t)-{\bf x}(t')\> |} \> . 
\label{S eff}
\ee
He then performed a variational approximation with a quadratic retarded 
trial action
\be
S_{\rm t} \E \frac{1}{2} \int_0^{\beta} dt \> {\bf \dot x}^2
+  \frac{1}{2} \int_0^{\beta} dt \, 
dt' \> f(t-t') \> \left [ \> {\bf x}(t) - {\bf x}(t') \> \right ]^2 \;\;\;,
\label{S trial}
\ee
choosing $ \> f_F(\sigma) \E  C \> \exp (- w \sigma) \> $,
where $C$ and $w$ are two variational parameters (the strength parameter
$C$ is usually written as $w (v^2-w^2)/4$ ).
This yields one of the best analytical approximations for the ground-state 
energy of the polaron for all values of the dimensionless coupling 
constant $\alpha$.
The best possible, rotationally invariant, quadratic trial action
is obtained by replacing the
exponential retardation by an arbitrary function $ f ( \sigma ) \> $
and was proposed by Adamowski {\it et al.} \cite{AGL} and 
Saitoh \cite{Sai}.   
Surprisingly for both small and strong coupling
this best isotropic quadratic approximation only yields very small 
improvements for the ground-state energy $E_0$
\bea
E_0 &\buildrel \alpha \to 0 \over \longrightarrow & - \alpha - a \, \alpha^2 
+ {\cal O}(\alpha^3) 
\label{eq: small alpha} \\
    &\buildrel \alpha \to \infty \over \longrightarrow& - \bar a \, \alpha^2 
+ {\cal O}(1) \>. 
\label{eq: large alpha}
\eea
One finds at small coupling $ \>a_F = 0.0123457, \> a_{\rm iso} = 
0.0125978\> $
whereas the exact value is $  \> a = 0.0159196 \> $ and for strong coupling
$ \> \bar a_F = \bar a_{\rm iso} = 0.106103  \> $ compared to 
$ \> \bar a = 0.108513 \> $.  
Probably discouraged by these analytical results
the best isotropic quadratic approximation has never been investigated 
systematically for a whole range of couplings, in particular for intermediate 
coupling.

It is the purpose of the present Letter to do this for the 
ground-state energy and the effective mass of the polaron and to point out 
that an easy, efficient
method exists to solve the non-linear variational equations.
Although numerical results have been reported in the literature
\cite{AGL,GLS}, it will turn out that they are unreliable and 
considerably {\it overestimate} the improvement on Feynman's approach. 
We also investigate the
question whether the inclusion of anisotropic terms in the quadratic trial
action leads to further improvements.
This work is an outgrow of recent attempts to generalize
Feynman's polaron approach to four-dimensional field theories [7-11] in 
the context of the worldline formalism \cite{worldline}. The 
nomenclature is the one used in Ref. \cite{WC1}. 

\vspace{1.4cm}
\noindent
{\bf 2.} We start with the expression for the ground-state energy using the 
quadratic trial action (\ref{S trial}):
\be
E_0 \E \frac{3}{2 \pi} \int_0^{\infty} dE \> \left [ \> \ln A(E) + 
\frac{1}{A(E)} - 1 \> \right ] \> - \> \frac{\alpha}{\sqrt{\pi}} 
\int_0^{\infty}
d \sigma \> \frac{e^{-\sigma}}{\mu(\sigma)} \> \equiv \> \Omega + V \> .
\label{E best}
\ee
Here 
\be 
\mu^2(\sigma) \E 
\frac{4}{\pi} \int_0^{\infty} dE \> \frac{\sin^2 ( E \sigma / 2 )}{E^2} 
 \> \frac{1}{A(E)}
\label{amu2 by A(E)}
\ee
is the ``pseudotime'' corresponding to the ``profile function'' $A(E)$
which is linked to to the retardation function through
\be
A(E) \E 1 + \frac{8}{E^2} \int_0^{\infty} d\sigma \> f(\sigma) 
\sin^2 \left ( \frac{E \sigma}{2} \right ) \> .
\label{prof by ret func}
\ee
With Feynman's choice one obtains the standard 
expression $A_F(E) = (v^2 + E^2)/(w^2 + E^2) \> $ for the profile 
function and the pseudotime can also be given analytically.
However, one can do better by not imposing a special form for the 
retardation function \cite{AGL,Sai}. Indeed, by varying Eq. (\ref{E best}) 
with respect
to $f(\sigma)$ or, equivalently, with respect to $A(E)$, one finds that 
the best variational profile function is determined by 
\be
A_{\rm iso}(E) \E 1  + \frac{4 \alpha}{3 \sqrt{\pi}} \> 
\int_0^{\infty} d\sigma \> \frac{e^{-\sigma}}{\mu^3_{\rm iso}(\sigma)} 
\frac{\sin^2 ( E \sigma/ 2 )}{E^2} \> .
\label{var eq for A(E)}
\ee
By comparing with Eq. (\ref{prof by ret func}) one sees that the 
retardation function has the form
\footnote{It should be kept in mind that for the ground-state energy 
we only consider the $\beta \to \infty$ limit which leads to the already 
simplified expression given in Eq. (\ref{S eff}).}
\be
f_{\rm iso}(\sigma) \E \frac{\alpha}{ 6 \sqrt{\pi}} \> 
\frac{e^{-\sigma}}{\mu^3_{\rm iso}(\sigma)} \> .
\label{var ret func} 
\ee
In particular, since for small $\sigma$ the pseudotime $\mu^2(\sigma)$
behaves like $\sigma$, we see that the retardation function 
$f_{\rm iso}(\sigma)$
has a $ \sigma^{-3/2} $-singularity at small relative times in marked 
contrast to Feynman's Ansatz. Equivalently, the profile function at 
large $E$ does not approach unity like $ 1/E^2$ as Feynman's 
parametrization suggests but slower, like 
$ \> A_{\rm iso}(E) \to 1  + \, 2 \sqrt 2 \alpha/(3 E^{3/2}) \> $ .
The wrong small-time behaviour of Feynman's Ansatz
is responsible for the awkward behaviour of his variational parameters
for $\alpha \to 0$: both $v$ and $w$ tend to 3 in this limit whereas 
one would have expected $w \to 1$ by comparison with the exact effective 
action (\ref{S eff}).

For a quantitative description at arbitrary $\alpha $
one has to solve the coupled non-linear
variational equations (\ref{var eq for A(E)}) and (\ref{amu2 by A(E)})
numerically. This can be done as follows \cite{WC2}:
one maps the infinite intervals to finite ones by substituting
$ \> E = \tan^2 \theta \> , \> \> \sigma = \tan^2 \psi \> $ 
and solves the variational equations on a grid of Gaussian points. In 
this way the required integrals both in the variational equations and in 
the evaluation
of the ground-state energy can be evaluated directly by Gauss-Legendre 
quadrature. The above choice of mapping is made to eliminate the integrable
but numerically slowly convergent square-root behaviour near $\sigma \to 0$
in the various $\sigma$-integrals.
The variational equations are then solved iteratively by starting
either from the perturbative values $ A(E) = 1, \> \mu^2(\sigma) = \sigma $
or from Feynman's parametrization using the known values of the variational
parameters \cite{LuRo}. For larger coupling the latter method gives faster
initial convergence as seen in Fig. 1. 
The convergence rate was monitored 
by calculating the maximal relative deviation between two iterations,
both for $A(E)$ and for
$\mu^2(\sigma)$ for all Gaussian points, i.e. for all 
discrete values $\theta_i$ and $\psi_i$. In the numerical calculations 
reported in Table 1 the allowed maximal relative deviation was set to 
$ 10^{-7}$
and reached after 7 - 70 iterations, depending on the coupling
constant (see Fig. 1). Furthermore, 
the $ [0,\pi/2] $-range for the 
variables $\theta, \psi$ was subdivided into $n_e$ intervals
and Gaussian integration with $n_g$ points was applied in the subintervals.
Typically values of $(n_e,n_g) = (6,72)$  were needed to obtain a 
ground-state energy accurate to six digits. It turned out that solving the 
non-linear variational equations by this method is numerically not more 
demanding than minimizing the energy functional with respect to the 
Feynman parameters $v$ and $w$.

\unitlength1mm
\begin{figure}[htb]
\begin{center}
\vspace{-2cm}
\mbox{\epsfysize=10cm\epsffile{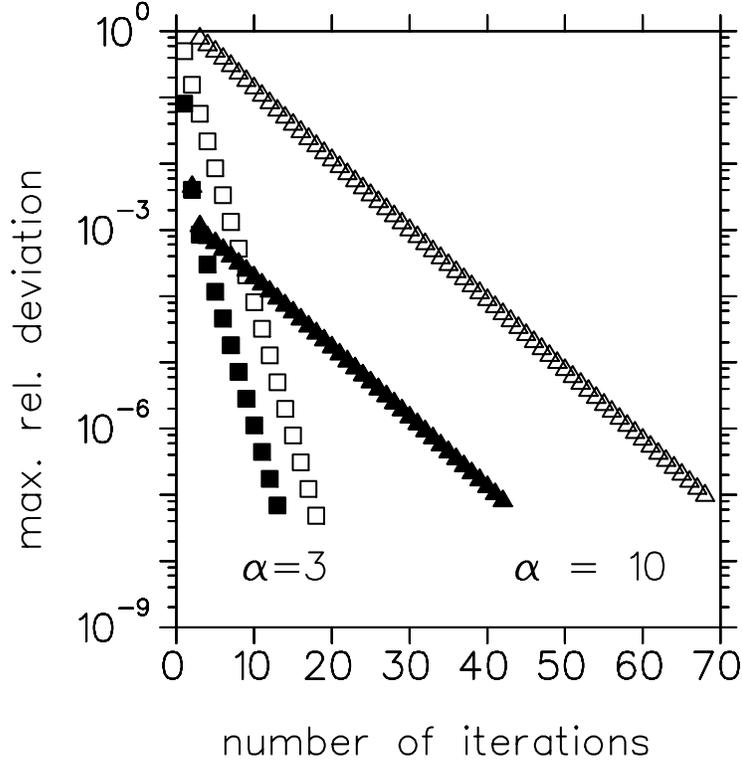}}
\end{center}
\caption{Convergence rate for solving the variational equations
(\ref{var eq for A(E)}, \ref{amu2 by A(E)}) iteratively at two different
values of the coupling constant. The open symbols denote the case when
the starting values are the free ones, the full symbols when Feynman's
parametrization is used for initialization.}
\label{fig:convergence}
\end{figure}

To check the numerical stability of the method the ground-state energy
was also calculated in a different way: the variational equations for
$A_{\rm iso}(E)$ can be used to re-express the kinetic term 
$ \> \Omega \> $ in terms of the potential $V$ (see the Appendix of 
Ref. \cite{WC2} where this is detailed
for the relativistic case):
\be 
\Omega_{\rm iso} \E \int_0^{\infty} d \sigma \> 
\frac{\delta V}{\delta \mu^2(\sigma)} \, \mu^4(\sigma) 
\frac{\partial}{\partial \sigma} \left ( \frac{\sigma}{\mu^2(\sigma)} 
\right)\Biggr|_{\mu^2 = \mu^2_{\rm iso}} \> .
\ee
This can be combined with the potential term and 
after elimination of the derivative of the pseudotime one obtains
the expression
\be
E_{\rm virial} \E - \frac{\alpha}{ \sqrt \pi} \int_0^{\infty} d \sigma \>
\left ( \frac{3}{2} - \sigma \right ) \, 
\frac{e^{-\sigma}}{\mu_{\rm iso} (\sigma)}   \> . 
\label{E virial}
\ee
This may be called ``virial energy'' since it also arises from
evaluating the virial theorem for two-time actions \cite{AlRo}.
In the numerical calculations an agreement in six digits between the
two expressions for the energy was demanded and obtained.

Table 1 collects the results for coupling constants from $\alpha = 1$
to $\alpha = 15$. Note that convergence is obtained in {\it all} cases
whereas in the relativistic case for coupling constants larger than a 
critical coupling the relative maximal deviations started to grow again
with increasing number of iterations \cite{WC2}. This was a signal for 
the instability of the scalar Wick-Cutkosky model whereas the polaron 
has a well-known strong coupling 
limit \cite{strong}. Also included is the effective mass of the polaron, 
defined by the expansion 
$E({\bf p}) = E_0 + {\bf p}^2/(2 m^{\star}) + \cdots \> $. 
It is given by \cite{GLS}
\be
m^{\star}_{\rm iso} \E 1 \> + \frac{\alpha}{3 \sqrt{\pi}} 
\int_0^{\infty} d\sigma \> 
\frac{\sigma^2}{\mu^3_{\rm iso} (\sigma)} e^{-\sigma}
\label{mstar best}
\ee
and coincides with the value of the profile function $A_{\rm iso}(E)$
at $E = 0$ (see Eq.~(\ref{var eq for A(E)})).

Being a variational calculation the values for the ground-state energy
should lie {\it below} Feynman's result which is non-optimal.
Table 1 shows that for small coupling the relative change
of the energy follows the behaviour 
$ \> (a_F - a_{\rm iso} ) \, \alpha = - 2.52 \cdot 10^{-4} \, \alpha \> $ 
which is expected 
from the small-coupling expansion (\ref{eq: small alpha}). But
also for intermediate coupling the best isotropic quadratic 
approximation gives only a slight improvement, maximally 0.15 \% ,
for the ground-state energy.

\begin{table}[htb]
\begin{center}
\begin{tabular}{|c||r|c|c||r|c|c|} \hline
         &             &       &        &          &       &   \\
$~~\alpha$~~ & $E_0^{\rm iso}~~ $\ \  & $ 10^3 \cdot \frac{E_0^{\rm iso}- E_0^F}{E_0^
F}$ 
& $\alpha^2 \cdot \left (E_0^{\rm iso}- E_0^F\right )$ &
$m^{\star}_{\rm iso}~~ $ &$10^3 \cdot \frac{m^{\star}_{\rm iso} - 
m^{\star}_F}{m^{\star}_F}$&
 $m^{\star}_{\rm iso} - m^{\star}_F$ \\ 
         &             &        &          &            &        &   \\
\hline
         &             &        &          &            &        &   \\
 1       & -1.013296   & 0.26   &           & 1.19615   & 0.53   &  \\
 2       & -2.056467   & 0.54   &           & 1.47515   & 2.21   &  \\
 3       & -3.135951   & 0.84   &           & 1.89862   & 5.12   &  \\
 4       & -4.261309   & 1.13   &           & 2.60234   & 8.93   &  \\
 5       & -5.447781   & 1.40   &           & 3.93259   & 12.1~~ &  \\
 6       & -6.72130~   & 1.55   &           & 6.9081~   & 10.2~~ &  \\
 7       & -8.12440~   & 1.44   &           & 14.416~~  & 1.52   &  \\
 8       & -9.70625~   & 1.12   &  -0.70    & 31.479~~  & -2.86  &  -0.09   \\
 9       & -11.49505~  & 0.81   &  -0.75    & 62.597~~~ & -2.46  &  -0.15   \\
10       & -13.4982~~  & 0.58   &  -0.78    &111.65~~~~ & -1.48  &  -0.17   \\
11       & -15.7163~~  & 0.41   &  -0.79    &182.96~~~~ & -0.90  &  -0.16  \\
12       & -18.1489~~  & 0.30   &  -0.79    &281.47~~~~ & -0.54  &  -0.15   \\
13       & -20.7954~~  & 0.23   &  -0.80    &412.65~~~~ & -0.32  &  -0.13   \\
14       & -23.6554~~  & 0.17   &  -0.81    &582.47~~~~ & -0.20  &  -0.11   \\
15       & -26.7285~~  & 0.13   &  -0.81    &797.4~~~~~~& -0.12  &  -0.10    \\ 
         &             &        &          &            &        &   \\ 
\hline
\end{tabular}
\end{center}
\vspace{-0.2cm}
\caption{The polaron ground state energy and effective mass in the best isotropic
quadratic 
approximation as a function of the coupling constant $\alpha$. The third 
and sixth column display the relative deviation from the results using
Feynman's parametrization (see Ref. \cite{LuRo} and Table \ref{tab:Feyn}), 
while the entries in the
fourth and last columns should approach a constant value in the large coupling 
limit (see text).} 
\label{tab:best}
\vspace{0.3cm} 
\end{table}

\noindent
In contrast, the numerical values for this approximation reported by 
Adamowski {\it et al.} claim to be
below the Feynman energy by as much as 0.75 \% at 
$\alpha = 11$.\footnote{The difference
might not seem much, however it should be remembered that Feynman's
energy is itself only a little more than 2 \% above the exact energy at
asymptotically large couplings; see Eq.~(\ref{eq: large alpha}).} That this
is incorrect and probably due to insufficient convergence and/or numerical
instability is shown in Fig. 2
where the present results for this specific
coupling constant are plotted as function of the number of Gaussian points.
It is clearly seen that with sufficient subdivisions of the
intervals both expressions 
(\ref{E best}) and (\ref{E virial}) for the ground-state energy
converge to the same value which is far away from the one given in 
Ref. \cite{AGL}.

\unitlength1mm
\begin{figure}[htb]
\begin{center}
\mbox{\epsfysize=10cm\epsffile{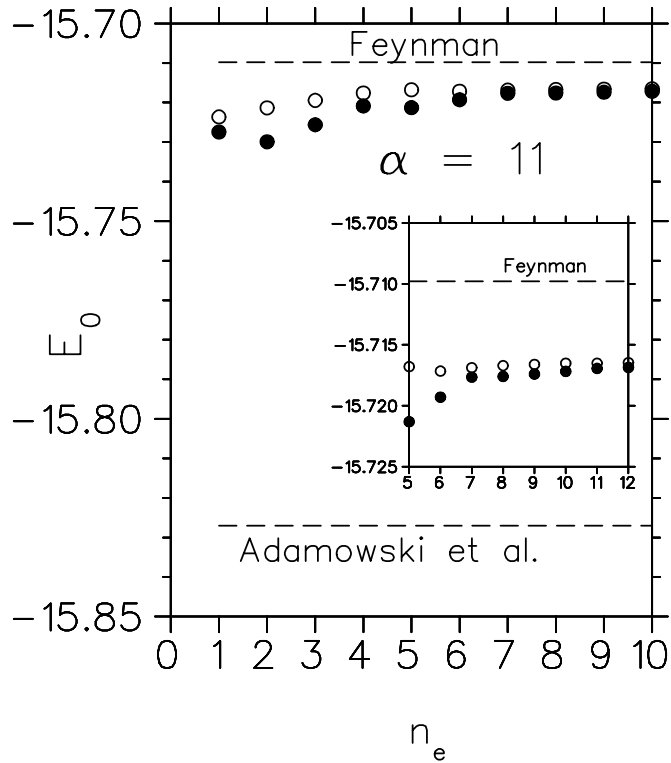}}
\end{center}
\vspace{-0.2cm}
\caption{Ground-state energy of the polaron 
for $\alpha = 11$ as a function of the number $n_e$ of subdivisions 
used in the 72-point Gauss-Legendre integrations. Open circles denote the 
energy evaluated from Eq. (\ref{E best}), full circles from the 
virial theorem (\ref{E virial}). The inset shows the convergence of
both numerical values in more detail.
The value using Feynman's  parametrization and the result 
reported by Adamowski {\it et al.} \cite{AGL} are shown by dashed lines.}
\label{fig:comparison}
\vspace{0.2cm}
\end{figure} 

An additional check on the accuracy of the numerical results is provided
by the large coupling limits of  $E_0^{\rm iso}$ and  $E_0^F$, which may be 
calculated analytically from Eq. (\ref{E best}).   A systematic expansion
around $\alpha \rightarrow \infty$ may be developed by making use of the
fact that as the coupling constant increases the pseudotime remains
constant over an increasing range of $\sigma$ (see Sec. 3.2 as well as the 
Appendix of Ref.~\cite{WC3}).  One finds, after some effort, that not
only the leading [${\cal O}(\alpha^2)$] term of $E_0^{\rm iso}$ and  $E_0^F$
is identical (see Eq. (\ref{eq: large alpha})), but also the first subleading 
term:
\be
E_0^{F,{\rm iso}} \> = \> -{1 \over 3 \pi} \alpha^2 \> - \> 3 \left (
{1 \over 4} + \log 2 \right ) \> + \> {\cal O}\left (\alpha^{-2}\right )
\;\;\;.
\ee
The numerical values in the fourth column of Table 1 indeed show
that the difference between $E_0^{\rm iso}$ and  $E_0^F$ is of order 
$\alpha^{-2}$.

For the effective mass the corrections to Feynman's result are somewhat
larger (for small $\alpha$ the relative change follows the perturbative
result 
$ \> 2 (a_{\rm iso} - a_F ) \, \alpha^2 = 5.04 \cdot 10^{-4} \, \alpha^2 \> 
$, 
and they are closer to the ones given in Ref. \cite{GLS}. For example,
at $\alpha = 5$ (their largest value) Gerlach {\it et al.} give 
$ m^{\star}_{\rm iso} = 3.940 $ whereas a similar increase in Gaussian 
points as
shown in Fig. 2 stabilizes the correct value at 
$ m^{\star}_{\rm iso} = 3.93259 $.
At larger couplings the convergence of the present method deteriorates 
somewhat (see Fig. 1) but up to  $\alpha = 15$ 
four to five significant figures can be given also for the
effective mass of Eq. (\ref{mstar best}). Surprisingly, the large coupling 
limits of 
both $m^{\star}_{\rm iso}$  and $m^{\star}_F$ again coincide to the first 
subleading term, i.e.,
\be
m^{\star}_{F,{\rm iso}} \> = \> {16 \over 81 \pi^2} \alpha^4
\> - \> {4 \over 3 \pi} \left( 1 + \log 4 \right ) \alpha^2
\> + \> {\cal O}\left (\alpha^0\right ) \> ,
\ee
the numerical values quoted in the
last column of Table 1 being consistent with this.
In the course of these calculations it 
turned out that the values of the effective mass
in Feynman's parametrization are very sensitive to the precise
numbers for the parameters $v, w$ and deviate somewhat from the ones given in 
Table II of Ref. \cite{LuRo}. We have re-calculated them by solving the 
variational equations for $v, w$ (instead of minimizing the
energy functional which is sufficient for the energy but not for the 
effective mass) and checked that they agree to a very high precision with the 
large-coupling expansion.
For convenience and further reference the new values are given in 
Table 2.

\begin{table}[thb]
\begin{center}
\begin{tabular}{|r||r|c||r|r||r|r|} \hline
         &           &           &              &   &           &      \\
$\alpha$ & $v$~~~    &   $w$     & $E_F$~~~~    &  $E_F$ (SC) & $m^{\star}_F$~~~~ &
$m^{\star}_F$ (SC) \\
 
         &           &           &              &               &           &      \\
  
\hline  
         &           &           &              &            &             &    \\   
 1       & ~3.10850  & 2.86958   & ~-1.0130308  &            & ~~1.19552   &    \\
 2       & ~3.24535  & 2.72564   & ~-2.0553560  &            & ~~1.47189   &     \\
 3       & ~3.42130  & 2.56031   & ~-3.1333335  &            & ~~1.88895   &     \\
 4       & ~3.66464  & 2.36792   & ~-4.2564809  &            & ~~2.57931   &     \\
 5       & ~4.03434  & 2.14002   & ~-5.4401445  &            & ~~3.88562   &     \\
 6       & ~4.66687  & 1.87363   & ~-6.710871~  &            & ~~6.83836   &     \\
 7       & ~5.80989  & 1.60365   & ~-8.112688~  &            & ~14.3941~~  &   \\
 8       & ~7.58682  & 1.40329   & ~-9.695371~  &            & ~31.5693~~  &   \\
 9       & ~9.85025  & 1.28230   & -11.485786~  &            & ~62.7515~~  &   \\
10       & 12.4749~  & 1.20918   & -13.490437~  & -13.489166 & 111.816~~~  & 111.150 
\\
11       & 15.4132~  & 1.16209   & -15.709808~  & -15.709373 & 183.125~~~  & 182.692 
\\
12       & 18.6483~  & 1.12988   & -18.143395~  & -18.143230 & 281.622~~~  & 281.327 
\\
13       & 22.1733~  & 1.10676   & -20.790681~  & -20.790613 & 412.782~~~  & 412.573 
\\
14       & 25.98515  & 1.08952   & -23.651278~  & -23.651248 & 582.584~~~  & 582.432 
\\
15       & 30.08224  & 1.076285  & -26.724904~  & -26.724900 & 797.498~~~  & 797.385 
\\ 
         &           &           &              &            &             &     \\ 
\hline
\end{tabular}
\end{center}
\vspace{-0.2cm}
\caption{Precise values for Feynman parameters, ground state energy and 
effective mass. The values in columns 5 and 7 are 
from the strong coupling (SC) expansions $ \> E_F(SC) = \sum_{i=0} e_i \, \alpha^{2-i
} 
\> \> , m^*_F(SC) =   \sum_{i=0} m_i \, \alpha^{4-i} \> $ for which we have 
calculated the following expansion coefficients: $ e_0 = -0.1061033 \> , e_2 = - 2.82
9442
\> , e_4 =  - 4.86387 \> , e_6 = -34.1953 \> , e_8 = 533.141 \> , e_{10} = 51525.2 \>
 ,
e_{12} = 6.61224 \cdot 10^6 \> , e_{14} = 9.53627 \cdot 10^8 \> $ and 
$ m_0 =  0.02001406 \> , m_2 = - 1.012775 \> , m_4 =  11.8558 \> , m_6 = 43.0986 \> $
. 
Most of these coefficients, except $e_{12}$ and  $e_{14}$, have also 
been obtained in Ref. \cite{SeSm}.}
\label{tab:Feyn}
\vspace{0.3cm}
\end{table}

\vspace{1cm}
\noindent

{\bf 3.} For a moving polaron Eq. (\ref{S trial}) is {\it not} the most
general quadratic trial action. Indeed, there is now a preferred direction
$ \hat p = {\bf p}/|{\bf p}| $ which may be used for constructing a trial
action which makes use of this directionality, i.e.,
\be
S_{\rm aniso} \E S_{\rm iso} - i ( \lambda' - 1) \, 
{\bf p} \cdot \int_0^{\beta} dt \> \dot {\bf x}(t)  
+ \frac{1}{2} \int_0^{\beta} dt \, dt' \> 
g(t - t') \, \Bigl ( \, \hat {\bf p} \cdot \left [ {\bf x}(t) -
{\bf x}(t') \right ] \, \Bigr )^2 \> . 
\label{S most gen quad}
\ee
This the most general quadratic trial action which is scalar and invariant 
under time and space translations as well as time-reversal. 
Time-translational invariance requires that the retardation functions in the
quadratic terms are functions of the relative time and allows only
a constant parameter (written as $\lambda' - 1$) in the linear term. 
Space-translational invariance leads to a dependence on co-ordinate
differences or derivatives
\footnote{Terms with derivatives may be converted into the form 
of Eq. (\ref{S most gen quad}) by suitable integrations by part.}. 
Finally, time-reversal invariance requires 
the linear term to be purely imaginary.
In Eq. (\ref{S most gen quad}) 
it has been written in such a way that it may be combined with 
the exponential in the Fourier transform of 
the partition function projected on momentum  $ {\bf p} $ \cite{AlRo}. 
For $ {\bf p} \ne 0 $ we thus have a new variational parameter 
$ \lambda' $ and {\it two} profile functions at our disposal 
\be
A(E) \> \longrightarrow \> A_{i j} (E) \E A_L(E) \, \hat p_i \hat p_j   + 
A_T(E) \, \left ( \delta_{ij}
-\hat p_i \hat p_j \right ) 
\hspace{1cm} i,j = 1, 2 ,3 \> .
\ee
Here the subscripts ``L'' and ``T'' denote longitudinal and transverse 
components
with respect to the direction of the polaron momentum.
Using the methods developed in Ref.~\cite{Adelaide} it is 
straightforward to calculate the various averages needed in 
the Feynman variational principle and to evaluate the limit
$\beta \to \infty$. One finds that the energy
\be
E_{\bf p}^{\rm aniso} \E \frac{1}{3} \Bigl ( \, 2 \, \Omega[A_T] + 
\Omega[A_L] \, 
\Bigr ) + V_{{\bf p}}[\mu_L^2,\mu_T^2] + 
\frac{{\bf p}^2}{2} \left ( 2 \lambda - \lambda^2 \right )
\label{E p}
\ee
is stationary 
\footnote{This is due to the use of a complex trial function and ``momentum
averaging'' but as in Ref. \cite{WC1} one can show that the energy for 
arbitrary momentum ${\bf p}$ is also minimal. The effective mass, however, 
is the derivative of the energy with respect to ${\bf p}^2$ at ${\bf p}=0$ 
and thereby has no minimum property anymore.}
under variation of the parameter 
$ \lambda = \lambda'/A_L(0) $ and the two profile functions $A_{L,T}(E) $
or, equivalently, the two pseudotimes $ \mu_{L,T}^2(\sigma) $.
The potential term is given by
\be
V_{{\bf p}} \E - \frac{\alpha}{\sqrt{\pi}} \int_0^{\infty} \! \! d\sigma \, 
e^{-\sigma} \! \int_0^1 \! \! dx \, \frac{\mu_L(\sigma)}{\mu_L^2(\sigma) 
+ \left [ \mu_T^2(\sigma) - \mu_L^2(\sigma) \right ] x^2} \, 
\exp \left  [ \frac{\lambda^2 {\bf p}^2 \sigma^2 
x^2}{2 \mu_L^2(\sigma) } \right ] \;\;\;.
\label{V p}
\ee
Clearly, the symmetric solution 
$A_T(E)=A_L(E)\equiv A_{\rm iso}(E)$ and hence
$\mu_T^2(\sigma)=\mu_L^2(\sigma)$,
emerges when ${\bf p}^2 = 0$ and hence, not surprisingly, 
$E_0^{\rm aniso}=E_0^{\rm iso}$.
A little 
less obvious is that the effective mass will, in fact, also remain
unchanged.  To see this, we expand the exponential in the potential to order
${\bf p}^2$:
\be
V_{{\bf p}}[A_T,A_L,\lambda]\> \equiv \>
V_0[A_T,A_L] \> + \> {\lambda^2 {\bf p}^2 \over 2} V_1[A_T,A_L] 
\> + \>{\cal O}\left ({\bf p}^4\right)\;\;\;.
\ee  
To leading order in the momentum
\be
 V_1[A_T,A_L]\> = \>V_1[A_{\rm iso}]\> + \>{\cal O}\left ({\bf p}^2\right)
\> = \>1-m^{\star}_{\rm iso}\> + \>{\cal O}\left ({\bf p}^2\right)
\label{eq: v1approx}
\ee
where Eq. (\ref{mstar best}) has been used.
On the other hand, $V_0$ (and  $\Omega$) contain an additional term of 
order ${\bf p}^2$  because of the implicit dependence of $A_{T,L}$
on the momentum. By writing 
$A_{T,L}(E) \equiv A_{\rm iso}(E) +  {\bf p}^2\> \Delta A_{T,L}(E)+ \ldots$
and expanding the functionals,
one can make 
make this remaining momentum dependence explicit. Hence one obtains for 
the energy
\bea
E_{{\bf p}}^{\rm aniso} \EA  \Omega[A_{\rm iso}] + V_0[A_{\rm iso}] + 
\> {\bf p}^2 \int _0^{\infty} dE \> \Biggl \{ \, \left [ \, \frac{2}{3} 
\Delta A_T(E) 
+  \frac{1}{3} \Delta A_L(E) \, \right ] \, {\delta \over \delta A(E)} 
 \Omega[A] \nonumber \\
 && + \sum_{i=L,T} \Delta A_i(E)  \, {\delta \over \delta A_i(E)} \, 
V_0[A_T,A_L] 
\, \Biggr \} _{A=A_{\rm iso}} +  {{\bf p}^2 \over 2} 
\left ( 2 \lambda - \lambda^2 + \lambda^2 V_1[A_{\rm iso}]\right)
\> + {\cal O}\left ({\bf p}^4\right) \nonumber \\
\EA E_0^{\rm iso} \> + \> {\bf p}^2 \int_0^{\infty} dE \>
{2  \Delta A_T(E)\> +\> \Delta A_L(E) \over 3} \>
\left[{\delta \over \delta A(E)} \Big(\Omega[A]+V_0[A]\Big)\right]
_{A=A_{\rm iso}}\> \nonumber \\
&& \hspace{5cm}
\> + \> {{\bf p}^2 \over 2} 
\left ( 2 \lambda - \lambda^2 + \lambda^2 V_1[A_{\rm iso}]\right)
\> + \>  {\cal O}\left ({\bf p}^4\right) \> .
\eea

\unitlength1mm
\begin{figure}[hbt]
\begin{center}
\mbox{\epsfysize=10cm\epsffile{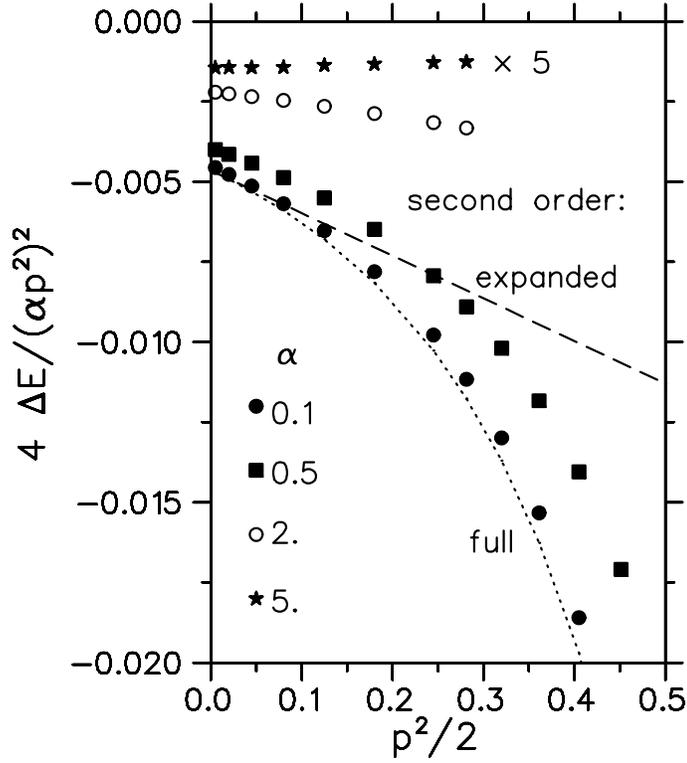}}
\end{center}
\caption{Energy difference between the variational approximation with 
the most general (anisotropic) and the isotropic trial action as function of 
the
coupling constant $\alpha$ and the polaron momentum ${\bf p}$. 
The numerical results at $\alpha = 5 $ have been multiplied by a factor 
$ 5 $ to
make them better visible in the graph.
The curves show the analytical perturbative ${\cal O}(\alpha^2) $ result: the 
dashed curve includes only the leading term at
small momenta (i.e. ${\cal O}({\bf p}^4)$, 
see Eq.~(\ref{diff an-iso}) ) while
the dotted line is the full 
second-order result without expansion in ${\bf p}^2$.}
\label{fig:aniso}
\vspace{0.7cm}
\end{figure} 

\noindent
The integral over $E$ vanishes for any $\Delta A_{T,L}(E)$ because
of the variational equation for $A_{\rm iso}(E)$.  On the other hand,
the variational equation for $\lambda$ yields $\lambda^{-1} = 1 - 
V_1[A_{\rm iso}]$
so that, upon substitution of Eq.~(\ref{eq: v1approx}), one obtains
\be
E_{{\bf p}}^{\rm aniso} \E E_0^{\rm iso} \> + \> {{\bf p}^2 
\over 2 m^{\star}_{\rm iso}}
\> + \>  {\cal O}\left ({\bf p}^4\right)\;\;\;.
\ee
It is only at higher orders in ${\bf p}^2$ that a difference between
$E_{{\bf p}}^{\rm aniso}$ and $E_{{\bf p}}^{\rm iso}$ can show up.
In fact, by solving the variational equations perturbatively, one sees
that there is a difference between these two energies at ${\cal O}(p^4)$:
\be
E_{{\bf p}}^{\rm aniso} -E_{{\bf p}}^{\rm iso} \E - \left ( 
\frac{{\bf p}^2}{2} 
\right )^2 \left[ 
\left({8 \over 225 \pi} - {1 \over 150}\right )\alpha^2 
\>+ \> {\cal O}\left ({\bf \alpha}^3\right)\right ]
\>+ \> {\cal O}\left ({\bf p}^6\right)\;\;\;.
\label{diff an-iso}
\ee
The anisotropic trial action has lowered the energy, as it should,
however note  that numerically the improvement is very small, the 
dimensionless
coefficient of the $\alpha^2 {\bf p}^4/4 $ term being only $ 0.004651 \> $.
Fig.~3 shows that for small couplings our numerical results
obtained by solving the anisotropic variational equations indeed follow the 
analytical prediction (\ref{diff an-iso}) rather well.

\vspace{1.4cm}
\noindent
{\bf 4.}
In summary, we have investigated the use of the most general quadratic 
trial actions, both isotropic and anisotropic,
in Feynman's variational approach to the polaron problem and found out that
they only lead to small numerical improvements compared to Feynman's 
original parametrization of the retardation function.  This is despite the
fact that the variational retardation functions have quite a different
small-time behaviour compared to Feynman's Ansatz, but is consistent with 
results from the large-$N$ expansion ~\cite{Smo},
where $N$ is the number of space dimensions (similar results have been 
reported in Refs. \cite{GaEf}).
Anisotropic terms in the trial action 
only become beneficial for a finite polaron momentum but not for the 
ground-state energy at ${\bf p} = 0$ and its derivative,
the effective mass. 

The small improvement over Feynman's results, even with arbitrary 
retardation functions, indicates that a quadratic trial action leaves 
out part of the correct physics of the polaron. 
This is most evident when the second-order expression for the ground-state 
energy is compared with the exact result.
However, this does not imply that the optimal quadratic trial actions,
both the isotropic   and the anisotropic one, are
useless. First, in four-dimensional
relativistic field theory the correct ultraviolet behaviour of the 
trial action is of much greater importance~\cite{WC2} than for the polaron.
Second, as shown in Ref.~\cite{WC3},
the correct analytic structure of the profile function $ A(E) $ is 
essential for describing scattering processes, where 
an analytic continuation to normal time is required. 
For these purposes Feynman's Ansatz is totally inadequate whereas
an improved parameterization
$ f_I(\sigma) \propto \exp(-w' \sigma)/\sigma^{3/2} \> $ 
captures the essence of the variational retardation function 
(\ref{var ret func}) and allows an easy analytic continuation. 
Third, extensions to fermionic theories like Quantum Electrodynamics 
require precise but subtle relations between fermionic and bosonic 
retardation
functions \cite{QED} which are only fulfilled by the 
variational solutions if the variational Ansatz is left sufficiently 
general.  And, finally, for a massive relativistic particle on its
gerlachmass-shell which corresponds to a moving 4-D polaron with
$p^2 = M^2 $, the 4-momentum is large and therefore one may expect 
substantial improvements in describing the physics of the system (for 
example, its instability) if one uses the most general, anisotropic 
trial action.

\vspace{3cm}
\noindent
{\bf Acknowledgement}: One of the authors (R. R.) would like to thank 
M. Smondyrev for helpful discussions and remarks and 
G. Ganbold for pointing out additional references.


\vspace{2cm}

\begin{thebibliography}{99}


\bibitem{GeLo} B. Gerlach and H. L\"owen, Rev. Mod. Phys. 63 (1991) 63.

\bibitem{MCM} T. K. Mitra, A. Chatterjee and S. Mukhopadhyay, Phys.
Rep. 153 (1987) 91.

\bibitem{Fey} R. P. Feynman, Phys. Rev. 97 (1955) 660.

\bibitem{AGL} J. Adamowski, B. Gerlach and H. Leschke, in: J. P. Antoine 
and E. Tirapegui (Eds.), Functional Integration - Theory and Application, 
Plenum, New York, 1980, p. 291 .

\bibitem{Sai} M. Saitoh, J. Phys. Soc. Jpn. 49 (1980) 878.

\bibitem{GLS} B. Gerlach, H. L\"owen and H. Schliffke, Phys. Rev. B36
(1987) 6320.

\bibitem{WC1} R. Rosenfelder and A. W. Schreiber, 
 Phys. Rev. D53 (1996) 3337.  
%%CITATION = NUCL-TH 9504002;%%

\bibitem{WC2} R. Rosenfelder and A. W. Schreiber,  
 Phys. Rev. D53 (1996) 3354.  
%%CITATION = NUCL-TH 9504005;%%

\bibitem{WC3} A. W. Schreiber,  R. Rosenfelder and C. Alexandrou,
 Int. J. Mod. Phys. E5 (1996) 681.
%%CITATION = NUCL-TH 9504023;%%

\bibitem{WC>3} A. W. Schreiber and R. Rosenfelder, 
Nucl. Phys. A601 (1996) 397; 
%%CITATION = NUCL-TH 9510032;%%
C. Alexandrou, R. Rosenfelder and A. W. 
Schreiber, Nucl. Phys. A628 (1998) 427; 
%%CITATION = NUCL-TH 9701036;%%
N. Fettes and R. Rosenfelder, 
Few-Body Syst. 24 (1998) 1 .
%%CITATION = FBSYE,24,1;%%

\bibitem{QED} C. Alexandrou, R. Rosenfelder and A. W. Schreiber, Phys. Rev. 
A59 (1999), 1762; Phys. Rev. D62 (2000) 085009 .
%%CITATION = HEP-TH 9809101;%%

\bibitem{worldline} M. J. Strassler, Nucl. Phys. B385 (1992) 145;
%%CITATION = HEP-PH 9205205;%%
M. Reuter, M. G. Schmidt and Ch. Schubert, Ann. Phys. (N. Y. ) 259 (1997) 
313 .
%%CITATION = HEP-TH 9610191;%%

\bibitem{LuRo} Y. Lu and R. Rosenfelder, Phys. Rev. B46 (1992) 5211 .
%%CITATION = PHRVA,B46,5211;%%

\bibitem{AlRo} C. Alexandrou and R. Rosenfelder, Phys. Rep. 215 (1992) 1.
%%CITATION = PRPLC,215,1;%%

\bibitem{strong} S. I. Pekar, "Untersuchungen zur Elektronentheorie der 
Kristalle", Akademie-Verlag, Berlin, 1954; S. Miyake, J. Phys. Soc. 
Jpn. 38 (1975) 181. 

\bibitem{SeSm} O. V. Selyugin and M. A. Smondyrev, Phys. Stat. Sol. (B) 155 
(1989) 155.
 
\bibitem{Adelaide} R. Rosenfelder, C. Alexandrou and A. W. Schreiber, in:
    A. W. Schreiber, A. G. Williams and A. W. Thomas (Eds.), 
    Proceedings of the Workshop on Methods in Non-Perturbative Field 
    Theory, Adelaide (Australia), 2 -13 February 1998, World Scientific, 
    Singapore, 1998, p. 163.

\bibitem{Smo} M. A. Smondyrev, Physica A171 (1991) 191; M. A. Smondyrev, in: 
H. Grabert, A. Inomata, L. Schulman and U. Weiss (Eds.), Proceedings 
of the Fourth International Conference on Path Integrals from meV to MeV:
Tutzing '92, World Scientific, Singapore, 1993, p. 190.

\bibitem{GaEf} G. Ganbold and G. V. Efimov, Phys. Rev. B50 (1994) 3733;
J. Phys. C10 (1998) 4845.

\end{thebibliography}
\end{document}